\documentclass[useAMS]{mn2e}
\usepackage{times}
\usepackage{arydshln}
\input{epsf}

\title[Inner disc radius in XTE J1817--330]
{X-ray irradiation in XTE J1817--330 and the inner radius of the truncated disc in the hard state}

\author[M. Gierli{\'n}ski, C. Done and K. Page]
{Marek
Gierli{\'n}ski$^{1}\thanks{E-mail:Marek.Gierlinski@durham.ac.uk}$,
Chris Done$^{1}$
and Kim Page$^{2}$\\
$^1$Department of Physics, University of Durham, South Road,
Durham DH1 3LE, UK\\
$^2$Department of Physics and Astronomy, University of Leicester, Leicester
LE1 7RH, UK\\
}

\date{Submitted to MNRAS}
\pagerange{\pageref{firstpage}--\pageref{lastpage}} \pubyear{2008}

\begin{document}

\topmargin = -0.5cm

\maketitle

\label{firstpage}

\begin{abstract}

The key aspect of the very successful truncated disc model for
the low/hard X-ray spectral state in black hole binaries is that
the geometrically thin disc recedes back from the last stable
orbit at the transition to this state. This has recently been
challenged by direct observations of the low/hard state disc from
CCD data. We reanalyze the {\it Swift} and {\it RXTE} campaign
covering the 2006 outburst of XTE J1817--330 and show that these
data actually strongly support the truncated disc model as the
transition spectra unambiguously show that the disc begins to
recede as the source leaves the disc dominated soft state. The
disc radius inferred for the proper low/hard state is less
clear-cut, but we show that the effect of irradiation from the
energetically dominant hot plasma leads to an underestimate of
the disc radius by a factor of 2--3 in this state. This may also
produce the soft excess reported in some hard-state spectra. The
inferred radius becomes still larger when the potential
difference in stress at the inner boundary, increased colour
temperature correction from incomplete thermalization of the
irradiation, and loss of observable disc photons from
Comptonization in the hot plasma are taken into account. We
conclude that the inner disc radius in XTE J1817--330 in the
low/hard spectral state is at least 6--8 times that seen in the
disc dominated high/soft state, and that recession of the inner
disc is the trigger for the soft--hard state transition, as
predicted by the truncated disc models.

\end{abstract}

\begin{keywords}

X-rays: binaries -- accretion, accretion discs

\end{keywords}

\section{Introduction}
\label{sec:introduction}

The accretion flow in black hole binaries (BHB) shows a distinct
transition at luminosities of $\sim$0.02--0.2 $L_{\rm Edd}$
(Eddington luminosity). Above this transition the spectra are
soft, peaking at $\sim$1~keV (in $\nu F_\nu$ representation) with
a quasi--thermal shape which is well described by models of a
geometrically thin, optically thick, cool disc. A series of such
disc-dominated spectra from a single object at different mass
accretion rates generally shows that the temperature and
luminosity of this component vary together as expected for a
constant inner radius, as predicted for a thin disc which extends
down to the last stable orbit around the black hole, and then
spirals quickly in towards the horizon (see e.g. the review by
Done, Gierli{\'n}ski \& Kubota 2007, hereafter DGK07).

However, below this transition luminosity the spectra are hard,
peaking at $\sim$100~keV, entirely unlike the predictions of a
standard disc model at these luminosities.  There is as yet no
clear consensus on the structure of the accretion flow in this
low/hard state, but one very attractive solution is that the
accretion disc makes a transition to an optically thin,
geometrically thick flow. While the detailed properties of such
flows depend on the cooling mechanisms assumed e.g. radiation
(Shapiro, Lightman \& Eardley 1976), advection (Narayan \& Yi
1994), convection (Abramowicz \& Igumenshchev 2001) and winds
(Blandford \& Begelman 1999) they are generically much hotter
than a thin disc solution at the same mass accretion rate as they
cannot efficiently cool by blackbody radiation (because they are
optically thin). These models typically have temperatures around
100~keV, so can match the observations (e.g. the review by
Narayan \& McClintock 2008 and references therein), and the
distinct nature of the hard and soft spectral states has an
obvious origin in the very different nature and geometry of the
accretion flow.

A key component of this `truncated disc' model is that the inner
disc is replaced by the hot flow. Thus the inner edge of the
cool, thin disc must recede back from the last stable orbit as
the source transitions from the disc dominated soft state to the
low/hard state. There are several key pieces of evidence
supporting this.  The spectra in the low/hard state can be
roughly described by a power law from 1 to 100~keV, with photon
spectral index of 1.5--2.  Superimposed on this are clear
features from Compton reflection from optically thick, cool
material. The solid angle subtended by the reflecting material
{\em correlates} with continuum shape, with harder spectra
showing less reflection and less associated fluorescent iron line
emission. This is easy to explain in the truncated disc models,
as the continuum shape is determined by the seed photon
luminosity from the disc intercepted by the hot flow, while
reflection is determined by the hot flow illuminating the disc.
As the disc moves outwards, this gives both fewer soft photons
from the disc to Compton cool the flow, and a smaller solid angle
of the reflecting material (e.g Poutanen, Krolik \& Ryde 1997;
DGK07).

The rapid variability seen in the low/hard state also supports a
model with a moving characteristic radius. All the major
characteristic frequencies (low-frequency break and low-frequency
quasi-periodic oscillation) seen in the broad-band power spectra
move, with higher frequencies (smaller characteristic radii)
being seen for softer spectra, as predicted by the truncated disc
model (e.g. Churazov, Gilfanov \& Revnivtsev 2001; DGK07). Even
the behaviour of the radio jet can be (phenomenologically)
incorporated into these models, with the hot inner flow providing
large scale height plasma and magnetic field close to the horizon
to power the jet. The collapse of the hot flow as the source
makes a transition to the disc dominated state triggers the
collapse of the jet (e.g. Fender, Belloni \& Gallo 2004; DGK07).

While these support the truncated disc model, they are indirect
measures of the inner radius of the cool, thin disc. But direct
observations of the emission from the disc itself are difficult
due to the low disc temperature of any disc (truncated or
otherwise) at low luminosities, and the problem is compounded by
(generally high) interstellar absorption and the relatively high
energy threshold of the most prolific BHB observatory, {\it Rossi
X-ray Timing Explorer} ({\it RXTE}). Nonetheless, the direct disc
emission was convincingly detected in {\it Hubble Space
Telescope}, {\it Chandra} and {\it BeppoSAX} observations of the
low interstellar absorption system XTE J1118+480. This transient
had peak luminosity over an order of magnitude below the
transition to the high/soft state so stayed in the low/hard state
during its outburst. The temperature and luminosity of the disc
emission clearly showed that it was truncated a few hundred
Schwarzschild radii from the black hole (Chaty et al. 2003).

Thus the disc is plainly truncated at very low $L/L_{\rm Edd}$,
but extends down to the last stable orbit in the high/soft state.
Therefore it must start to recede at some point, and the
transition is the obvious place. However, Miller, Homan \&
Miniutti (2006a) and Rykoff et al. (2007, hereafter R07) have
claimed that observations of direct disc emission in the low/hard
state at luminosities close to the transition show that the disc
still extends down to the last stable orbit after the transition,
requiring another (as yet unknown) physical mechanism for the
transition. Here we re-examine the data of R07 from XTE
J1817--330, as this is the only CCD based (and hence low energy
instrument threshold) campaign following a BHB outburst with
reasonable sampling, and the low interstellar absorption towards
this object means that the soft X-ray disc component can be well
constrained. Startlingly, the simple disc fits show that the
inner disc recedes during the transition, consistent with the
truncated disc picture derived above (see also Cabanac et al., in
preparation). However, the apparent radius then {\em decreases}
as the source enters the low/hard state, back to a value
consistent with that seen in the high/soft state (see also R07).
However, we show that the apparent radius in the low/hard state
can be underestimated by a large factor through the effects of
X-ray irradiation (which also changes the colour temperature
correction from that expected in the high/soft state), together
with the changing stress condition at the inner boundary and
suppression of the observed seed photons by Compton scattering.
Thus the low/hard state in XTE J1817--330 is also consistent with
a truncated disc, and this physical mechanism can also explain
the apparently small radii for the inner disc seen in other
objects (e.g. Miller et al. 2006a). However, irrespective of the
detailed models used for the low/hard state, the transition
spectra {\em always} show that the disc inner radius is larger
than that in the high/soft state, giving strong confirmation that
disc truncation is the physical origin of the soft--hard state
transition.

\section{Data reduction}
\label{sec:data}

{\it RXTE} observed XTE J1817--330 from 2006 January 27 to August
2. We reduced these data using {\sc ftools} version 6.4. We
extracted energy spectra from detector 2 (top layer only) of the
Proportional Counter Array (PCA), and added 1 per cent systematic
error in each channel. We use these data for spectral fitting in
3--20 keV band. We also extracted power density spectra (PDS)
from all available detectors in full PCA energy band (2--60 keV)
and 1/256--64 Hz frequency band.

\begin{figure}
\begin{center}
\leavevmode \epsfxsize=8cm \epsfbox{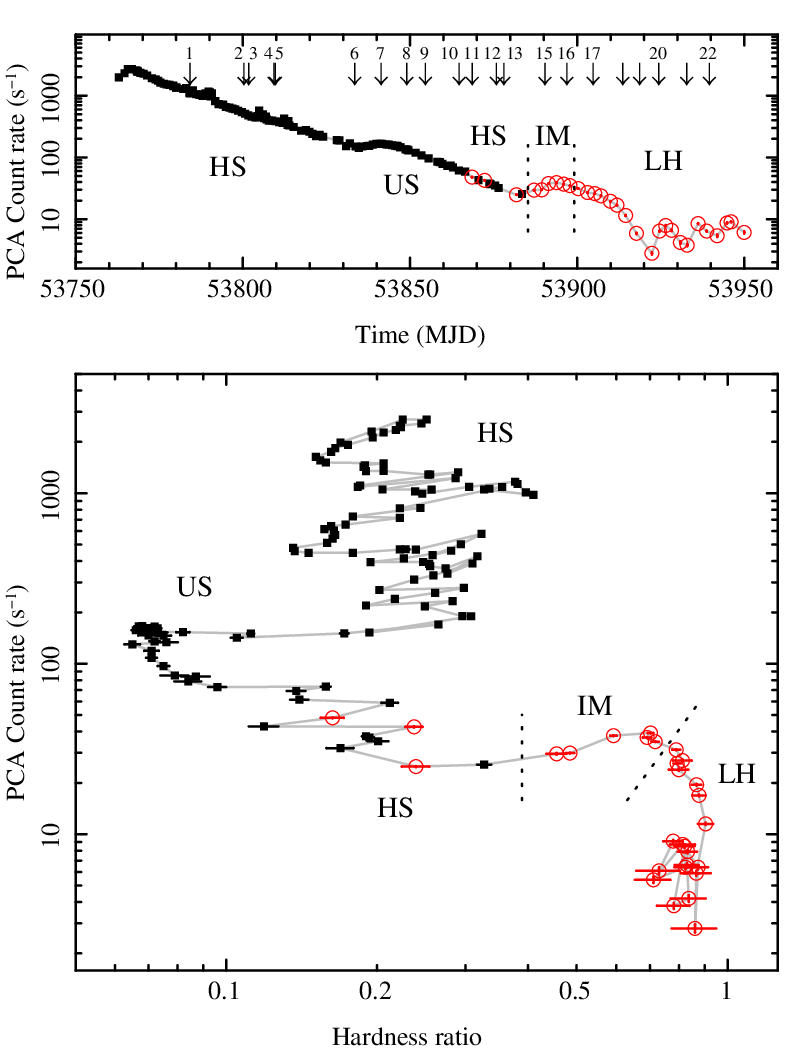}
\end{center}
\caption{Light curve (upper panel) and hardness-intensity diagram for
2006 outburst of XTE J1817--330 from {\it RXTE} PCA data, detector 2
only. Dotted lines separate spectral states: HS - high/soft, IM -
intermediate and LH - low/hard. US denotes ultrasoft state, an extreme
version of the high/soft state.  Filled symbols represent PCA
observations with well established disc parameters. Open red circles
show observations where either the disc was not statistically
significant in the PCA data or the fractional error on its
normalization was greater then 50 per cent. The arrows in the upper
panel show times of {\it Swift} observations. The numbers above arrows
show observation numbers, following R07 (only data with significant disc,
see Tables \ref{tab:simple} and \ref{tab:thir}).} \label{fig:hid}
\end{figure}

\begin{figure}
\begin{center}
\leavevmode  \epsfxsize=8cm \epsfbox{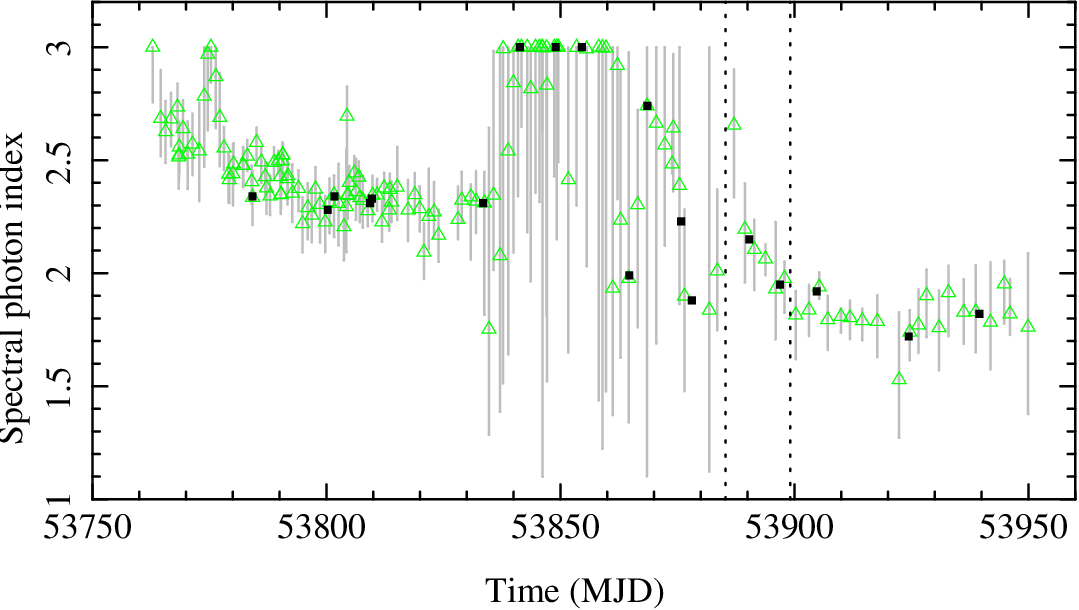}
\end{center}
\caption{Evolution of the X-ray photon spectral index, $\Gamma$,
during the outburst. Open green triangles show the best-fitting
values from the {\it RXTE} PCA data. Filled black squares
represent interpolated $\Gamma$ used for {\it Swift} fits. Dotted
lines represent onsets of the intermediate and hard states (see
Fig. \ref{fig:hid}). The interpolated values of $\Gamma$ are shown in Table \ref{tab:simple}}
\label{fig:gamma}
\end{figure}

The source was very bright at the start of the outburst, so the
{\it Swift} X-ray Telescope (XRT) was in Windowed Timing mode.
However, even this was piled up above 100 counts s$^{-1}$ so we
extracted these spectra similarly to R07 by excluding the inner,
piled up region as determined from where the ratio of grade 0 to
grade 0--2 events becomes constant. This was a 7 pixel radius
region for the brightest spectra 1--10, then 2 pixels for
spectrum 11 and none for spectra 12--17 (observation numbers as
in table 1 of R07). The XRT was switched into Photon Counting
mode as the source was so much dimmer, though again pileup became
an issue above 0.6 counts s$^{-1}$ so the exclusion region was 5,
3 and 4 pixels for spectra 18, 19--21 and 22 respectively.

\section{Results}

\subsection{Spectral states} \label{sec:states}

Fig. \ref{fig:hid} shows the light curve and hardness-intensity
diagram (HID) of all 150 PCA observations, where hardness is the
ratio of count rate in the 6.3--10.5 to 3.8--6.3~ keV energy
bands (e.g. Fender et al. 2004). In order to better understand
spectral evolution of XTE J1817--330 during its outburst we
fitted all the PCA data with a simple model, consisting of disc
and Comptonization. The disc is modelled by standard {\sc diskbb}
model, while the high-energy tail by thermal Comptonization code
{\sc thcomp}\footnote{publicly available as a local {\tt xspec}
model at
http://heasarc.gsfc.nasa.gov/docs/xanadu/xspec/models/nthcomp.html}
(Zdziarski, Johnson \& Magdziarz 1996), with photon index
constrained to be less than 3. We fix the electron temperature at
50 keV, outside the bandpass of the spectral fits, and the
absorbing column at $1.2\times10^{21}$ cm$^{-2}$ (R07). We also
add a Gaussian line and smeared edge to approximately account for
the effects of reflection. We obtain good fits ($\chi^2/\nu <
1.25$) for all data sets.

Fig. \ref{fig:gamma} shows (open green triangles) the evolution
of the photon spectral index of Comptonization. Using this
information, together with the HID, and also by visually
inspecting all energy and power spectra, we identify four
distinct spectral states.  The upper branch on the HID with
hardness ratio $\sim 0.2$ is characterized by strong disc
emission, a weak high-energy tail and low rms variability,
typical of the high/soft state behaviour.  There is a short
excursion to higher hardness ratios at a count rate around 1000
s$^{-1}$ (MJD $\sim$~53790), taking the system to the edge of a
very high state, as also shown by the stronger variability
including a transient QPO. The source then drops to very low
hardness ratios ($\le 0.1$), sometimes termed an ultrasoft state
but this is simply an extreme case of the soft state, with a low
temperature disc and a very weak high-energy tail.

The focus of this paper is the transition to the hard state in
the later part of the outburst.  The onset of the transition is
marked by an increasing fraction of Comptonization giving
increasing hardness ratio, and increasing rms variability. The
transition is rather smooth so it cannot be pinpointed very
accurately. We use a hardness ratio of $0.4$ (and low count rate
$<$100 s$^{-1}$ to distinguish from the very high state) to mark
the beginning of the transition to an intermediate state, and the
point where the spectral index becomes less than 2 as the
beginning of the low/hard state. We use these boundaries to
identify the X-ray spectral states for {\it Swift} data. Thus
observations 1--13 are in the soft state, 15 and 16 are
intermediate spectra during the transition into the hard state,
while later observations are in the hard state (see table 1 in
R07).

\subsection{Disc radius from a simple disc model} \label{sec:simple_disc}

\begin{table}
\begin{tabular}{cccccccccc}
\hline
Obs & $kT_{\rm in}$ & $N_{\rm disc}$ & $\Gamma$ & $\chi^2_\nu$/d.o.f.\\
& (keV) & ($\times10^{3}$) \\
\hline
 1  & $0.89\pm0.01$              & $2.5\pm0.1$          & (2.34) & 588.2/539 \\
 2  & $0.796\pm0.006$            & $2.4\pm0.1$          & (2.28) & 606.3/527 \\
 3  & $0.779\pm0.007$            & $2.6\pm0.1$          & (2.34) & 519.0/494 \\
 4  & $0.701\pm0.007$            & $2.8\pm0.1$          & (2.31) & 534.7/487 \\
 5  & $0.71\pm0.01$              & $2.7\pm0.1$          & (2.33) & 477.2/451 \\
 6  & $0.59\pm0.01$              & $4.3\pm0.3$          & (2.31) & 331.0/316 \\
 7  & $0.62\pm0.01$              & $3.3_{-0.1}^{+0.2}$  & (3.00) & 338.1/361 \\
 8  & $0.583\pm0.015$            & $3.5\pm0.2$          & (3.00) & 358.3/290 \\
 9  & $0.592_{-0.007}^{+0.008}$  & $3.2\pm0.1$          & (3.00) & 344.9/339 \\
10  & $0.516\pm0.009$            & $3.7_{-0.2}^{+0.3}$  & (1.99) & 310.4/261 \\
11  & $0.504\pm0.005$            & $4.3\pm0.2$          & (2.74) & 421.5/337 \\
12  & $0.489\pm0.003$            & $4.2\pm0.1$          & (2.23) & 588.2/362 \\
13  & $0.478\pm0.003$            & $3.7\pm0.1$          & (1.88) & 672.3/360 \\
\hdashline[1pt/3pt]
15  & $0.392\pm0.004$            & $4.7\pm0.2$          & (2.15) & 432.9/367 \\
16  & $0.294\pm0.003$            & $10.1\pm0.5$         & (1.95) & 474.2/396 \\
\hdashline[1pt/3pt]
17  & $0.225_{-0.002}^{+0.004}$  & $18\pm1.5$           & (1.92) & 393.0/307 \\
20  & $0.18\pm0.03$              & $3_{-2}^{+4}$        & (1.72) &  40.8/40  \\
22  & $0.185_{-0.014}^{+0.017}$  & $6_{-2}^{+3}$        & (1.82) &  66.3/64  \\
\hline
\end{tabular}

\caption{Fit results of the simple disc model. $T_{\rm in}$ is
the disc temperature at the inner radius, $N_{\rm disc}$ is the
disc normalization, $\Gamma$ is the fixed Comptonization spectral
index interpolated from {\it RXTE} data (see also Fig.
\ref{fig:gamma}). Dotted horizontal lines separate the spectral
states (HS above, IM in the middle, LH at the bottom) in the same
way as dotted lines in the figures.} \label{tab:simple}
\end{table}

\begin{figure}
\begin{center}
\leavevmode  \epsfxsize=8cm \epsfbox{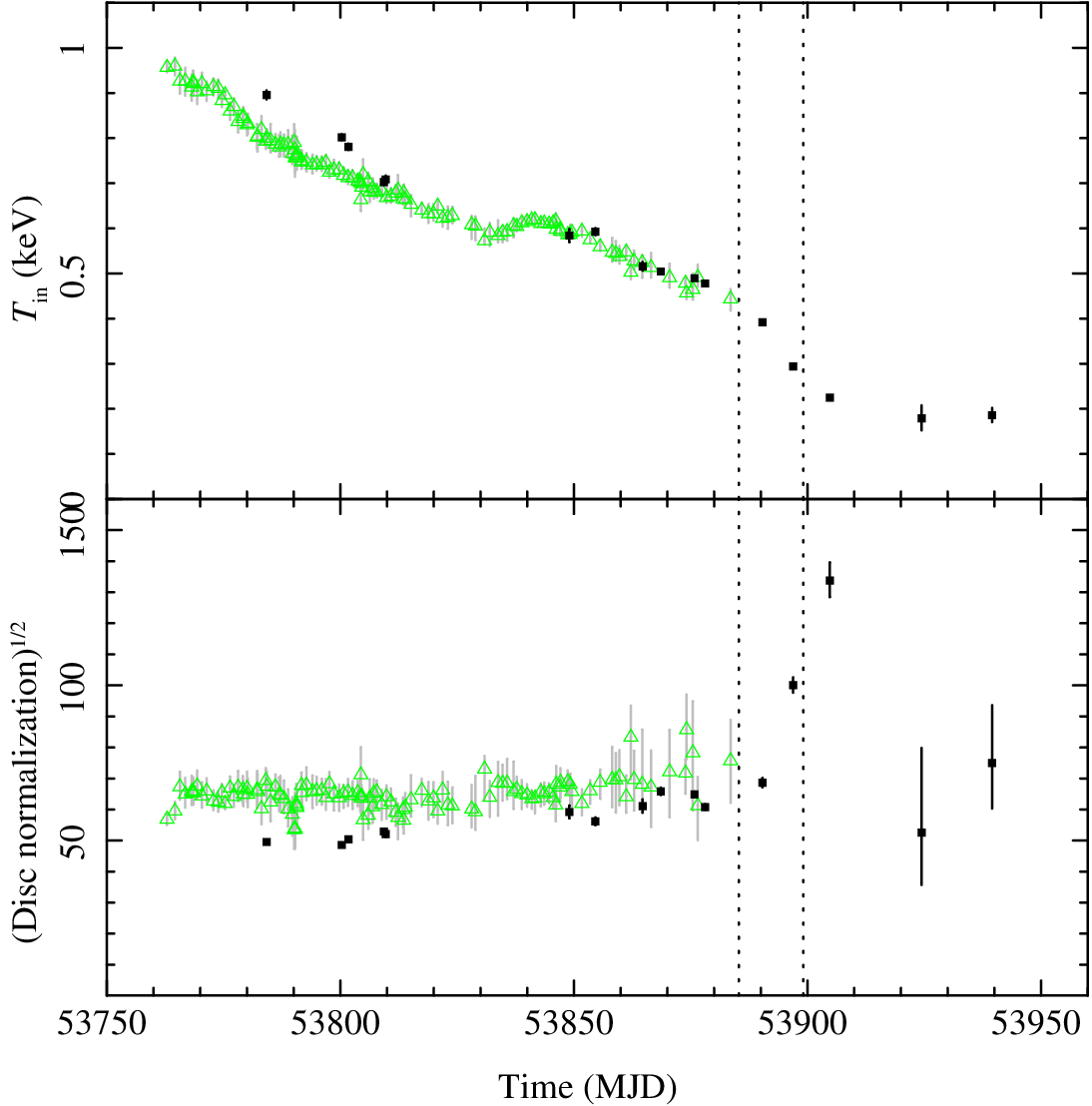}
\end{center}
\caption{Temperature and square root of normalization
(proportional to the inner disc radius) of the disc component.
The green open triangles show the {\it RXTE} PCA data; only fit
results with statistically significant disc and fractional error
on its normalization less then 50 per cent are shown, hence no
PCA data after $\sim$MJD 53875, where errors become very large.
The black filled squares show the {\it Swift} XRT observations
fitted with a standard disc model. Dotted
lines represent onsets of the intermediate and hard states (see
Fig. \ref{fig:hid}).} \label{fig:temprad}
\end{figure}

\begin{figure}
\begin{center}
\leavevmode  \epsfxsize=8cm \epsfbox{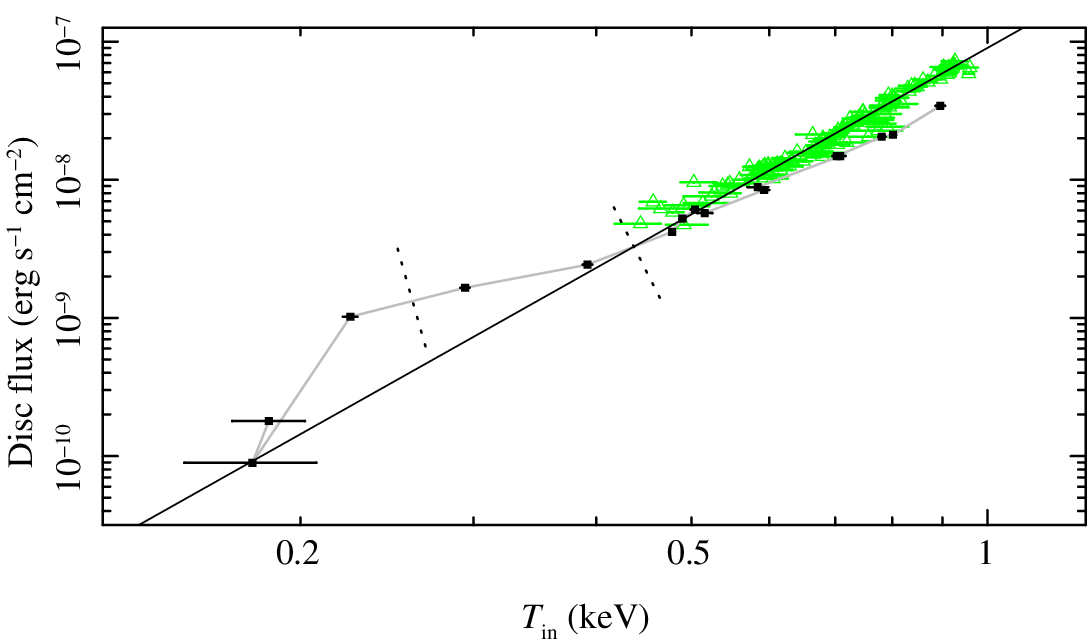}
\end{center}

\caption{Disc flux as a function of disc temperature. The meaning
of various symbols is the same as in Fig. \ref{fig:temprad}. The
dotted lines mark state transition, as in Fig. \ref{fig:hid}: HS
is on the right, IS in the middle and LH state on the left. The
straight line represents $F \propto T_{\rm in}^4$ relation.}
\label{fig:tempflux}
\end{figure}

The results from the simple disc model are shown in Table
\ref{tab:simple}. Fig. \ref{fig:temprad} shows the evolution of
the disc parameters (temperature and radius) throughout the
outburst as derived from the simple fits. We only include {\it
RXTE} spectra for which the disc properties can be well
constrained (we rejected the data where the uncertainty of
normalization is more than 50 per cent of normalization itself).
This selects {\em only} high state data, and excludes {\em all}
the intermediate and low state data which are the focus of this
paper. The inner disc temperature drops from about 0.95 to
0.4~keV, throughout the soft states but the disc inner radius
(proportional to square root of {\sc diskbb} normalization)
remains remarkably constant. This is equivalent to saying that
the  disc flux, $F$ and temperature, $T_{\rm in}$ closely follow
the expected thermal relationship $F \propto T_{\rm in}^4$ from a
constant emitting area, as shown in Fig. \ref{fig:tempflux}.

However, the lower limit of 3~keV on the low energy bandpass of
the PCA mean that these data cannot constrain the disc properties
of the soft-to-hard transition and the hard state. Lower bandpass
instruments (generally CCD's) are required, but these are often
not easy to schedule for extensive monitoring campaigns. {\it
Swift} carries perhaps the only CCD instrument for which the
required sampling can be feasibly obtained, and the data set of
R07 on XTE J1817--330 is outstanding in this regard, especially
as this object has low Galactic absorption.

Ideally, the {\it Swift} 0.35--10~keV spectra could be fit together
with quasi-simultaneous {\it RXTE} 3--20~keV data, but there is a
small but significant change in spectral shape between the data sets
which points to minor differences in calibration (probably enhanced to
uncertainties in pileup correction at high luminosities). Instead, we
fit only the {\it Swift} data, but fix the spectral index of the
Comptonization to that seen in harder X-ray PCA data.  As the {\it
Swift} and {\it RXTE} data are not quite simultaneous we interpolate
between the spectral indices measured from adjacent PCA observations.
This should give a good estimate as Fig.  \ref{fig:gamma} shows that
the spectral index generally varies on timescales of a few days,
rather longer than the typical offset of less than a day between the
{\it RXTE} and {\it Swift} observations.  Fixing the spectral index
reduces the number of free parameters in XRT fits, so helped to
constrain the disc temperature and normalization.  We also removed the
reflection features from the model, as they are not required to fit
the {\it Swift} data.

Most of the XRT spectra are well fit by this model, typically
giving reduced $\chi^2_\nu \sim 1$. The only exceptions are
observations 12 and 13, where $\chi^2_\nu$ = 1.62 and 1.87,
respectively. This is not due to the fixed Comptonization
spectral index, as allowing this to be free leads to a very
different type of fit. The spectrum is then dominated by a very
low temperature/steep spectral index Comptonization component, in
disagreement with the PCA data. This behaviour is typical of
fitting a simple {\tt diskbb} model to more sophisticated (hence
hopefully more realistic) disc spectra (Done \& Davis 2008) as
relativistic effects broaden the spectrum from that predicted by
{\tt diskbb}, and the shape is made more complex by atomic
features imprinted by radiative transfer through the vertical
structure of the disc (e.g. Ebisawa, Mitsuda \& Hanawa 1991;
Zhang, Cui \& Chen 1997; Li et al. 2005; Davis et al. 2005).
Observations 12 and 13 are the ones with highest signal-to-noise
(the source is brighter in earlier observations, but the pileup
limits mean that the extracted count rate is smaller), so are the
ones where this effect is most evident.

The results of these fits are shown in Figs. \ref{fig:temprad}
and \ref{fig:tempflux} in black filled squares.
The disc is not significantly
detected in the XRT data sets 18, 19 and 21 (due mostly to poor
signal--to--noise as these are short exposures) so we exclude these
from further analysis. There is a fairly
good agreement between the disc derived from the PCA and XRT
results in the soft state. Only the first 5 XRT observations (MJD
53784--53809) have slightly higher disc temperatures than
corresponding PCA data, which again may be due to small
differences in calibration between the PCA and XRT instruments,
probably enhanced by uncertainties in pileup correction at these
high count rates, and/or the different weighting of the {\tt
diskbb} model fits to the different band passes (Done \& Davis
2008).


The key aspect of the {\it Swift} data is that it extends the disc
constraints into the spectral transition and hard state
(Figs. \ref{fig:temprad} and \ref{fig:tempflux}). These data clearly
show that the inferred disc radius {\em increases} by a factor $\sim$2
during the transition (XRT observations 15, 16 and 17). The same
effect can be seen in tables 3 and 4 of R07, in clear confirmation of
the predictions of the truncated disc model.  However, the apparent
disc inner radius then decreases in the hard state proper, back to a
value which is consistent with the untruncated disc of the soft state
(R07).

\subsection{Irradiated disc model} \label{sec:irradiated_model}

\begin{figure}
\begin{center}
\leavevmode  \epsfxsize=8.5cm \epsfbox{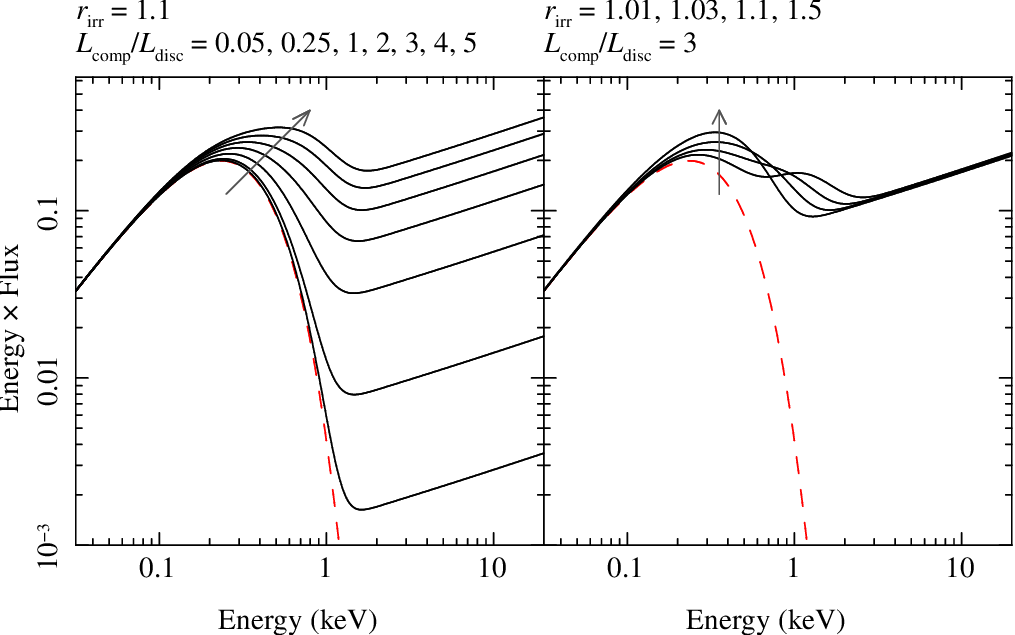}
\end{center}
\caption{Model of the Compton-irradiated disc as a function of
Compton-to-disc ratio, $L_{\rm c}/L_{\rm d}$ (left panel) and
irradiation radius, $r_{\rm irr}$ (right panel). The model was
calculated for $kT_{\rm in}$ = 0.1 keV, $\Omega/2\pi$ = 0.3,
$\Gamma$ = 1.7 and $kT_e$ = 100 keV. The grey arrows show the
direction of increasing $L_{\rm c}/L_{\rm d}$ (left panel) and
$r_{\rm irr}$ (right panel). The dashed line shows disc
with no irradiation.} \label{fig:model_diskir}
\end{figure}

\begin{figure}
\begin{center}
\leavevmode  \epsfxsize=8.5cm \epsfbox{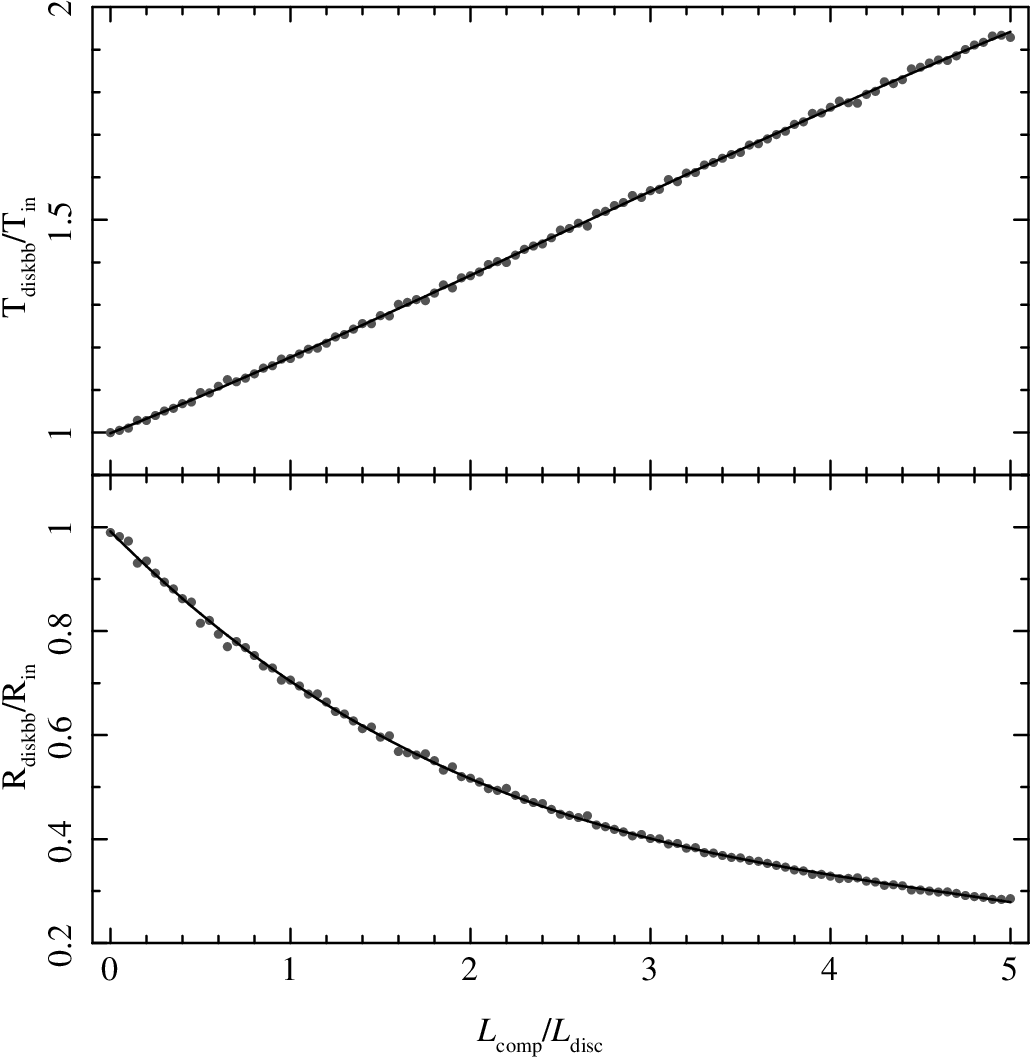}
\end{center}
\caption{Effect of irradiation on the disc measurements. The figure
shows the standard, non-irradiated disc parameters, when it is fitted
to a set of simulated data based on the irradiated disc.  When there
is no Comptonization ($L_{\rm c}/L_{\rm d} = 0$) both models provide
the same results. However, with an increasing fraction of
Comptonization and increasing irradiation, the disc becomes hotter
(see Fig. \ref{fig:model_diskir}). Fitting an irradiated disc with the
non-irradiated model gives higher temperature and smaller inner
radius. The gray points in the background show the actual fit results,
while the black lines are third-order polynomial fits to these
data. The data were simulated for $kT_{\rm in} = 0.1$ keV,
$\Omega/2\pi = 0.3$, $\Gamma$ = 1.7, $kT_e$ = 100 keV and $r_{\rm irr}
= 1.1$.}
\label{fig:model_ratios}
\end{figure}

The continuum model we used to fit XRT data in Section
\ref{sec:simple_disc} consists of the two standard components
expected in black hole spectra, namely the disc and its
Comptonization by energetic electrons. They are linked to each
other as the seed photons for Comptonization come from the disc,
so any change in disc temperature affects the low-energy shape of
Comptonization.  However, the converse is not included: the
simple models assume that Comptonization does not influence the
disc.

This is plainly an approximation, as the Comptonized component
illuminates the disc. Some fraction of this illuminating spectrum
produces the reflected continuum and iron line, while the
remaining, non-reflected, fraction is reprocessed and re-emitted
as quasi-thermal emission and adds to the observed disc emission
(e.g. Malzac, Dumount \& Mouchet 2005). Where the disc is
dominant energetically, the reprocessed emission from
illumination from the Compton tail is only a small fraction of
the intrinsic disc flux, so the observed disc spectrum consists
mostly of its intrinsic emission. However, once the Compton tail
carries a large fraction of the bolometric luminosity (as in the
intermediate and especially hard states) the reprocessed emission
can add substantially to the observed disc flux, affecting the
inferred disc temperature, luminosity and hence radius.

We make a simple model of an X-ray illuminated disc. The fraction
of Comptonized emission illuminating the disc is set by the solid
angle of the reflector, $\Omega/2\pi$.  The luminosity of the
reprocessed emission is then $L_{\rm rep} = \Omega/2\pi (1-a)
L_{\rm c}$, where $L_{\rm c}$ is the luminosity in the
Comptonized emission and $a$ is the energy and angle integrated
reflection albedo. This illumination adds to the intrinsic
(gravitational) energy disc luminosity, $L_{\rm d}$. Hence, the
total observed disc luminosity is the sum of intrinsic and
reprocessed luminosities, $L_{\rm d}+L_{\rm rep}$. We note that
illumination alone underestimates the total heating expected in
the truncated disc model, as the hot flow is in thermal contact
with the disc, so there is also conductive flux from electrons
(Meyer \& Meyer Hofmeister 1994; R{\'o}{\.z}a{\'n}ska \& Czerny
2000) and ions (Dullemond \& Spruit 2005) which can also heat the
disc.

Our model is designed to reproduce the standard {\tt diskbb}
spectrum in the limits where the illuminating flux is negligible,
so the intrinsic disc flux radial dependence is $r^{-3}$ (i.e.
has no stress free inner boundary correction), with unilluminated
disc luminosity $L_{\rm d}$. A simple order-of-magnitude
calculation shows that the reprocessed flux {\em dominates} the
observed disc flux, i.e. $L_{\rm rep} > L_{\rm d}$, if $L_{\rm
c}/L_{\rm d} \ga 10$, at which point the observed ratio between
the Compton tail and total disc emission is $L_{\rm c}/(L_{\rm
d}+L_{\rm rep}) \sim 5$. This holds for standard low/hard state
parameters of $\Omega/2\pi=0.3$ and $a=0.3$ (see e.g. Poutanen et
al. 1997; Gilfanov, Churazov \& Revnivtsev 1999; Ibragimov et al.
2005)

The temperature structure of the illuminated disc is set by the
(fourth root of the) sum of fluxes from the intrinsic and
reprocessed emission at each radius in the disc. However, the
radial dependence of the illumination which gives rise to the
reprocessed flux is less easy to define than that for the
intrinsic disc emission, as it depends on the relative geometry
of the disc and Comptonization region. Motivated by the truncated
disc scenario as sketched in Esin, McClintock \& Narayan (1997)
and DGK07, we consider especially the overlap region between the
disc and hot flow which is required at the transition and in the
bright low/hard states (Poutanen et al. 1997). Hence we assume
that the inner disc embedded in the hot flow is uniformly
illuminated by the hot flow, and that the illumination of the
rest of the disc is negligible. We scale this illumination
radius, $r_{\rm irr}$, in units of the inner disc radius.

We encode this model in {\sc xspec} using {\sc thcomp} for the
Comptonized emission. The model parameters are ratio of
luminosities in Comptonization and unilluminated disc, $L_{\rm
c}/L_{\rm d}$, and the irradiation radius $r_{\rm irr}$. There
are also Comptonization parameters, $\Gamma$, $kT_e$, and
$\Omega/2\pi$ in addition to the unilluminated disc inner
temperature, $T_{\rm in}$ and disc normalization proportional to
the square root of the inner radius, $R_{\rm
in}^{1/2}$.\footnote{this model, {\tt diskir}, is publicly
available on the {\sc xspec} web page
http://heasarc.gsfc.nasa.gov/software/xspec}

Fig. \ref{fig:model_diskir}a shows a sequence of spectra of the
irradiated disc model for increasing $L_{\rm c}/L_{\rm d}$ for
typical low/hard state parameters and $r_{\rm irr}$ = 1.1
(Poutanen et al. 1997). The intrinsic disc remains the same
throughout the sequence, i.e.  there is the same accretion rate
through a disc of constant radius. We also show the effect of
changing the size of the irradiated region, $r_{\rm irr}$, in
Fig. \ref{fig:model_diskir}b. Obviously, the smaller this is, the
higher the resulting temperature and vice versa. However, the
amount of overlap also sets the Comptonized continuum shape, so
this is {\em not} a free parameter (see Poutanen et al. 1997).

The disc spectrum is largely unaffected for $L_{\rm c}/L_{\rm d}
< 1$. However, hard state spectra have $L_{\rm c}/L_{\rm d} >1$
where our models predict that the soft X-ray part of the spectrum
deviates significantly from the original disc shape. There is a
rise at soft energies resembling a disc spectrum, but it is
significantly hotter then the underlying disc. If this is fit by
a simple {\sc diskbb} model, the resulting temperature is higher
and normalization (or radius) lower.

We quantify this effect by simulating a sequence of 100 spectra
with $L_{\rm c}/L_{\rm d}$ varying from 0 to 5 through the XRT
response.  We fit the simulated spectra by the same simple
non-irradiated model of disc and Comptonization we used in
Section \ref{sec:simple_disc}, and the resulting inferred
temperature and inner radius are shown in Fig.
\ref{fig:model_ratios}. We stress again that the underlying disc
in the simulated data has constant inner radius and mass
accretion rate, yet its inferred temperature can be up to a
factor $\sim$2 higher in the hard state. More importantly, the
measured, apparent inner radius coming out from spectral fits is
{\em underestimated} by up to a factor  $\sim$3.  This is crucial for
the truncated disc model!

The factor by which the radius is underestimated depends on the
strength of the irradiation, so depends also on the particular
choice of $\Omega/2\pi$ and $r_{\rm irr}$ as well as on $L_{\rm
c}/L_{\rm d}$. Increasing $\Omega/2\pi$ to 1 (as opposed to 0.3
in Fig. \ref{fig:model_ratios}) increases the radius a few times
in the limit of large $L_{\rm c}/L_{\rm d}$, while $r_{\rm
irr}=1.5$ (as opposed to 1.1) increases the radius by a factor
$\sim$1.2--1.6. However, we stress again that these are not all
independent factors. In a full model, the changing overlap
geometry between the hot flow and cool disc, $r_{\rm irr}$, sets
the solid angle which determines both the amount of reflection,
$\Omega/2\pi$ and the spectral index of the Comptonization. The
reprocessed flux is then determined by the reflection albedo (so
feeds back into the calculation of spectral index as it increases
the seed photon flux) and this is set by a combination of the
ionization state of the disc (the disc is more reflective for
high ionization) and the spectral index (high energy photons
cannot be reflected elastically due to Compton downscattering so
$a<0.3$ for hard spectra while $a\to 1$ for soft spectra
reflected from a highly ionized disc). Such models with full
coupled energetics (e.g. Malzac 2001; Malzac et al. 2005) are
beyond the scope of this paper but we hope to address these in
future work. Instead, here we simply set the irradiation
parameters to those appropriate for a low/hard state with
$\Gamma\sim 1.7$ (i.e. $\Omega/2\pi = 0.3$, $a=0.3$, $r_{\rm irr}
= 1.1$: Poutanen et al. 1997). These are not appropriate for the
soft state, but given that the reprocessed emission is negligible
in these spectra then this has little effect on the fits.

\subsection{Spectral fits with the irradiated disc model} \label{sec:irradiated_fits}

\begin{table}
\begin{tabular}{ccccccccccc}
\hline
Obs & $kT_{\rm in}$ & $N_{\rm disc}$ & $L_{\rm c}/L_{\rm d}$ & $\chi^2_\nu$/d.o.f.\\
& (keV) & ($\times10^{3}$) \\
\hline
 1 & $0.88\pm0.01$              & $2.8\pm0.1$            &  $0.12\pm0.03$           & 588.1/539 \\
 2 & $0.786\pm0.007$            & $2.6\pm0.1$            &  $0.08\pm0.01$           & 606.6/527 \\
 3 & $0.771\pm0.008$            & $2.7\pm0.1$            &  $0.06\pm0.02$           & 518.8/494 \\
 4 & $0.682\pm0.008$            & $3.3\pm0.1$            &  $0.18\pm0.02$           & 531.9/487 \\
 5 & $0.69\pm0.01$              & $3.2\pm0.2$            &  $0.18\pm0.02$           & 476.3/451 \\
 6 & $0.58\pm0.01$              & $4.7_{-0.3}^{+0.4}$    &  $0.1\pm0.02$            & 330.2/316 \\
 7 & $0.61\pm0.01$              & $3.7_{-0.2}^{+0.3}$    &  $0.07\pm0.02$           & 338.4/361 \\
 8 & $0.57\pm0.02$              & $4\pm0.4$              &  $0.08\pm0.03$           & 358.6/290 \\
 9 & $0.589\pm0.009$            & $3.3\pm0.2$            &  $0.02\pm0.01$           & 345.3/339 \\
10 & $0.51\pm0.01$              & $4\pm0.3$              &  $0.07\pm0.02$           & 309.4/261 \\
11 & $0.498\pm0.006$            & $4.7\pm0.2$            &  $0.06\pm0.01$           & 421.6/337 \\
12 & $0.484\pm0.004$            & $4.4\pm0.1$            &  $0.05\pm0.01$           & 584.4/362 \\
13 & $0.472\pm0.003$            & $3.9\pm0.1$            &  $0.08\pm0.01$           & 659.1/360 \\
\hdashline[1pt/3pt]
15 & $0.372\pm0.005$            & $6.2\pm0.3$            &  $0.36\pm0.02$           & 417.2/367 \\
16 & $0.267_{-0.004}^{+0.003}$  & $15.7_{-0.8}^{+0.9}$   &  $0.68\pm0.02$           & 430.6/396 \\
\hdashline[1pt/3pt]
17 & $0.199\pm0.004$            & $31\pm3$               &  $0.79\pm0.03$           & 373.5/307 \\
20 & $0.104_{-0.014}^{+0.017}$  & $27_{-13}^{+26}$       &  $3.4_{-0.8}^{+1.5}$     &  40.1/40  \\
22 & $0.148_{-0.013}^{+0.015}$  & $16_{-6}^{+9}$         &  $1.2\pm0.2$             &  65.8/64  \\
\hline
\end{tabular}

\caption{Fit results of the irradiated disc model. $T_{\rm in}$
is the intrinsic disc temperature at the inner radius, $N_{\rm
disc}$ is the disc normalization, $L_{\rm c}/L_{\rm d}$ is the
ratio of luminosity in the Comptonized component to that in the
non-irradiated disc (i.e. intrinsic disc emission). Dotted
horizontal lines separate the spectral states in the same way as
dotted lines in the figures.} \label{tab:thir}
\end{table}

\begin{figure}
\begin{center}
\leavevmode  \epsfxsize=8.5cm \epsfbox{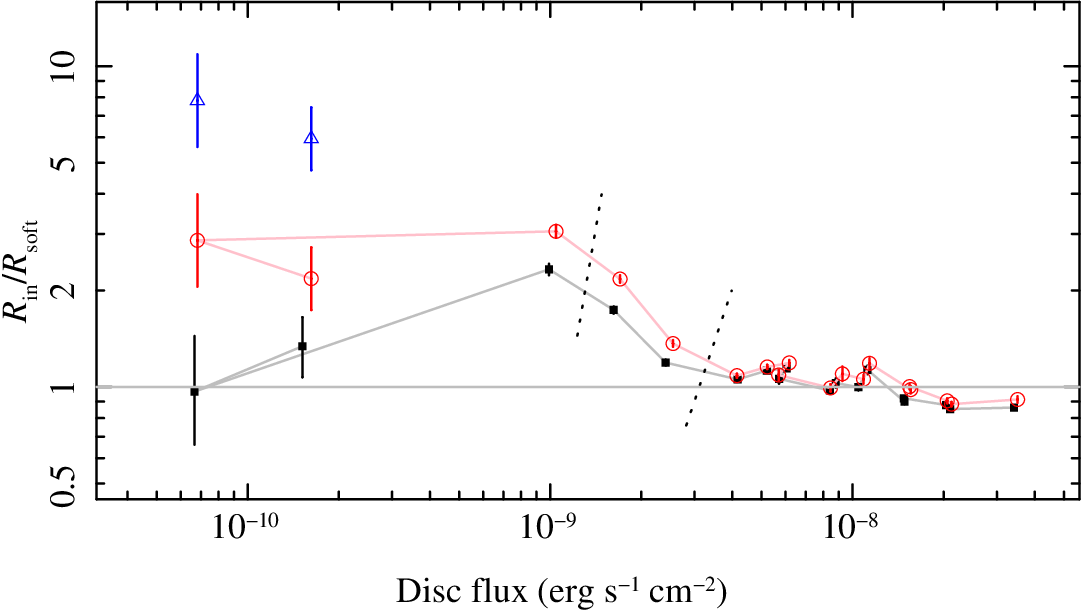}
\end{center}

\caption{The inner disc radius, normalized to the mean radius in
the soft state, as a function of the disc flux. Black filled
squares represent the simple non-irradiated model ({\sc diskbb});
red circles show the Compton irradiated model; blue triangles
refer to the irradiated model with stress free boundary condition
(low/hard state only, observations 20 and 22). The dotted lines
mark state transition, as in Fig. \ref{fig:hid}: HS is on the
right, IS in the middle and LH state on the left. }
\label{fig:flux_radius}
\end{figure}

The results from fitting the data with our irradiated disc model
are shown in Table \ref{tab:thir}. The disc inner radius as a
function of the disc flux is shown in Fig.  \ref{fig:flux_radius}
in open red circles. Black filled squares show the results from
non-irradiated {\sc diskbb} model, for comparison. As expected,
in the soft state, where contribution from Comptonization, and
irradiation, is weak, there is not much difference between the
two models. The soft-to-hard state transition, marked by a dotted
line begins when the disc flux drops below $\sim3\times10^{-9}$
erg s$^{-1}$ cm$^{-2}$. Even with the non-irradiated disc model
the inner disc radius increases by a factor 2 during the
transition, while with the irradiated model it increases by
factor 3, quantitatively but not qualitatively different.
However, the two models give quite different results in the hard
state, where the disc flux drops to $\sim10^{-10}$ erg s$^{-1}$
cm$^{-2}$ (the corresponding bolometric flux is then about 4
times larger). While the non-irradiated model yields the inner
disc radius comparable to that in the soft state, the irradiated
model gives $R_{\rm in}$ about 2--3 times larger, and hence
consistent with continuous truncation from the transition
onwards.

\begin{figure}
\begin{center}
\leavevmode  \epsfxsize=8.5cm \epsfbox{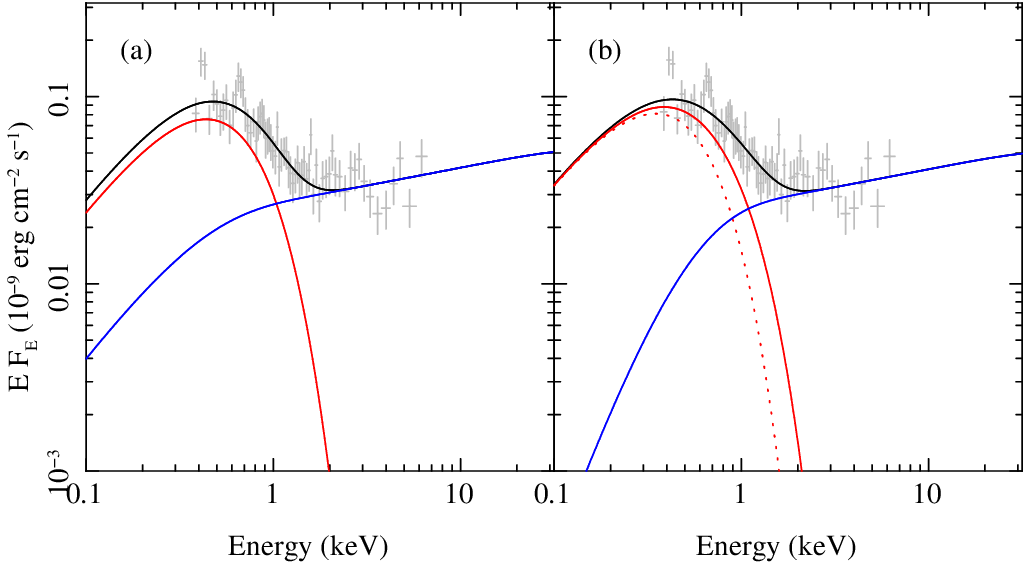}
\end{center}

\caption{Unfolded and unabsorbed X-ray spectra of hard-state
observation 22 (MJD 53939). The red curve (peaking at $\sim$0.4
keV) shows the disc, the blue curve (power law at higher
energies) shows Comptonization, while the black curve corresponds
to the sum of these. (a) Model with a standard non-irradiated
disc. (b) Model with the irradiated disc; the dotted line shows
the intrinsic emission of the disc from the gravitational
heating. The solid red line shows the total disc emission,
including contribution from Comptonization. In this model the
seed photons for Comptonization are assumed to be from the
irradiated section of the disc only, so they have higher
temperature and the low-energy cutoff in Comptonization is at
higher energies in compare to the standard model.}
\label{fig:spec}
\end{figure}

In Fig. \ref{fig:spec} we show the how the irradiated disc model
differs from the standard, non-irradiated disc. In addition to
the intrinsic emission from the disc from gravitational heating
(dotted line in panel $b$), there is contribution from
irradiation, which makes the disc hotter. When the same spectrum
is fitted by the standard disc model (plus Comptonization), the
disc temperature is overestimated and its inner radius
underestimated.

The error bars in Fig. \ref{fig:flux_radius} are quite large and
the recession of the disc in the hard state is not particularly
significant. However, there are other aspects of the accretion
disc structure which are different between this and the soft
state. Firstly, there is a stress-free inner boundary condition
on a thin disc which extends down to the last stable orbit (e.g.
Shakura \& Sunyaev 1973), as supported by observations (e.g.
DGK07). However, when the disc truncates at larger radii, then
the stress should be continuous at the inner edge. Thus a direct
comparison of disc spectra between the soft and hard states is
not appropriate even without irradiation since they do not have
the same stress condition an their inner boundary. The difference
between a stressed and unstressed disc equates to a changes in
inner radius by a factor $\sim$~2.7, hence to a truncation radius
in the hard state of 6--8 times that of the soft state. This
corresponds to 40 or 50 gravitational radii.

Yet another issue which may lead to underestimation of the disc
inner radius is Compton scattering. The hot plasma, which Compton
scatters seed photons from the accretion disc, has optical depth
of order unity. If this plasma lies in our line of sight to the
disc then the observed disc emission is suppressed by a factor
$\sim \exp(-\tau)$ i.e. or order 2--3 (Kubota \& Done 2004, Done
\& Kubota 2006). The total effect of this depends on geometry,
but for the truncated disc/hot inner flow the seed photons are
predominantly from the overlap region, so those on the far side
of the black hole are strongly suppressed, especially in high
inclination systems, while those on the near side are not. Thus
this can make up to a factor 2 decrease in disc flux i.e. a
factor $\sqrt{2}$ decrease in apparent disc inner radius
(Makishima et al. 2008).

The irradiation (by definition) heats the top layer of the disc.
This is unlikely to thermalize in the same way as the intrinsic
gravitational energy release, which is dissipated deep within the
disc. The irradiation flux most probably forms a moderately
Comptonized spectrum, i.e. has a larger colour temperature
correction than for the mostly thermalized soft state emission.
This effect alone would give rise to an decrease in apparent disc
radius by a factor $(f_{\rm col, hard}/f_{\rm col,
soft})^2\sim$~1.5 if the colour temperature correction changes
from $f_{\rm col, soft}\sim1.8$ in the soft state, to $f_{\rm
col, hard}\sim 2.2$ for all the flux to be dissipated in the disc
photosphere (Done \& Davis 2008, in preparation) in the hard
state.

Where irradiation dominates the energetics, the resulting spectrum may
be more complex than can be described by a single colour temperature
corrected blackbody or disc blackbody. Davis et al. (2005) give two
examples of spectra where half the energy is dissipated in the
photosphere, showing clearly that the spectrum is much better
described by Comptonization. This is very interesting as there is
growing evidence that there is an additional component in the hard
state spectra. This component, a `soft excess' of temperature higher
than the disc, is required in addition to disc and high temperature
Comptonization in some hard state spectra. It has been reported
several times in Cyg X-1 (Ebisawa et al. 1996; Di Salvo et al. 2001;
Ibragimov et al.  2005; Makishima et al. 2008) and in GRO J1655--40
(Takahashi et al. 2007). There is as yet no clear explanation for this
component (see the list of possibilities in Di Salvo et al.  2001),
but disc irradiation provides a physically plausible association. This
is testable as it predicts that the variability of the reprocessed
emission should follow that of the illuminating flux, with lag and
smearing typical of the light travel time across the inner disc. This
should also correlate with the variability of the iron line and
reflected continuum, as the same irradiating flux produces both
reprocessed and reflected emission.

A more complex spectrum for the soft emission may also
potentially explain the apparent extremely broad iron lines which
are sometimes inferred for the hard state. GX 339--4 is the best
example of this, where the line derived for a low/hard state
using simple continuum models requires a disc down to the last
stable orbit of a rapidly spinning Kerr black hole, completely
inconsistent with a truncated disc (Miller et al. 2006b).
Incomplete thermalization of the reprocessed disc emission from
irradiation means the spectrum extends up to higher energies than
predicted by a blackbody disc, which could affect the red wing of
the line. We will test this in a future paper, but here we note
that such broad lines are now the only challenge to the truncated
disc models for the hard state, whereas there are multiple pieces
of evidence which support them (e.g. DGK07).

\section{Conclusions} \label{sec:conclusions}

In this paper we demonstrate that the same data which are used by
R07 to rule out the truncated disc model actually strongly
support it. The transition spectra unambiguously show that the
disc begins to recede as the source leaves the disc dominated
soft state (Fig. \ref{fig:flux_radius}). The radius derived in
the hard state is less robust, but we show that irradiation from
the energetically dominant hot plasma can be crucial. The
reprocessed flux from the hard X-ray illumination can dominate
the intrinsic disc emission, leading to an underestimate of the
disc radius by factors of 2--3 (Fig. \ref{fig:model_ratios}).
Incorporating irradiation into the disc models gives hard state
radii which are consistent with the disc being truncated (Fig.
\ref{fig:model_diskir}). The inferred radius recedes still
further when the potential difference in stress at the inner
boundary, increased colour temperature correction from
irradiation, and loss of observable disc photons from
Comptonization in the hot plasma are taken into account. We
conclude that the inner disc radius in XTE J1817--330 in the hard
spectral state is at least 6--8 times that seen in the disc
dominated high/soft state, and that recession of the inner disc
is the trigger for the soft--hard state transition, as predicted
by the truncated disc models (Esin et al. 1997).

\section*{Acknowledgements}

We thank the anonymous referee for their helpful comments. CD and
MG acknowledge support through a PPARC senior fellowship, and
Polish MNiSW grant NN203065933, respectively. We thank Eli Rykoff
and Omer Blaes for discussions and encouragement.


\label{lastpage}

\end{document}